\documentclass[preprint,pra,showpacs]{revtex4-1}
\usepackage{graphicx, subfigure}
\usepackage{amsmath}
\usepackage{amsfonts}
\usepackage{tikz}
\newcommand\krho{k_{\rho}}
\newcommand\intkrho{\int_0^\infty d\krho}
\newcommand\kz{k_z}

\newcommand\xhat{\mathbf{\hat{x}}}
\newcommand\yhat{\mathbf{\hat{y}}}
\newcommand\rhat{\mathbf{\hat{r}}}
\newcommand\lhat{\mathbf{\hat{l}}}
\newcommand\zhat{\mathbf{\hat{z}}}

\newcommand\phihat{\mathbf{\hat{\phi}}}
\newcommand\rhohat{\mathbf{\hat{\rho}}}

\newcommand\Cnkrho{\mathbf{C}_{n\krho}}
\newcommand\Dnkrho{\mathbf{D}_{n\krho}}
\newcommand\Mr{\mathbf{M}(\mathbf{r})}
\newcommand\Nr{\mathbf{N}(\mathbf{r})}
\newcommand\Mreta{\mathbf{M}_\nu}
\newcommand\Nreta{\mathbf{N}_\nu}
\newcommand\Aplus{\mathbf{A}^+_\nu} 
\newcommand\Aminus{\mathbf{A}^-_\nu} 
\newcommand\Aplusbar{\mathbf{A}^+_{\bar{\nu}}} 
\newcommand\Aminusbar{\mathbf{A}^-_{\bar{\nu}}} 
\newcommand\Aplusdagger{{\mathbf{A}^+_\nu}^\dagger} 
\newcommand\Aminusdagger{{\mathbf{A}^-_\nu}^\dagger}

\newcommand\pws{\widehat{\textbf{s}} }
\newcommand\pwp{\widehat{\textbf{p}}}
\newcommand\pwplus{\widehat{\textbf{e}}_+ }
\newcommand\pwminus{\widehat{\textbf{e}}_- }
\newcommand\pwsthetaphi[2]{\widehat{\textbf{s}}_{#1,#2}}
\newcommand\pwpthetaphi[2]{\widehat{\textbf{p}}_{#1,#2}}

\newcommand\smallkrho{\frac{\krho}{k}\rightarrow 0}
\newcommand\cnkrho{c_{n\krho}}
\newcommand\dnkrho{d_{n\krho}}

\newcommand\shat{\widehat{\mathbf{s} }}
\newcommand\phat{\widehat{\mathbf{p} }}
\newcommand\phik{\phi_k}
\newcommand\thetak{\theta_k}
\newcommand\plane{\exp(i \textbf{k} \cdot \textbf{r})}

\newcommand\Ankrho{\mathbf{A}_{n\krho}}
\newcommand\Bnkrho{\mathbf{B}_{n\krho}}

\newcommand\rotation{\mathbf{R} (\thetak, \phik)}

\newcommand\alphak{\alpha(\thetak, \phik)}
\newcommand\betak{\beta(\thetak, \phik)}
\newcommand\cp{\cos \phik}
\newcommand\ct{\cos \thetak}
\newcommand\spp{\sin \phik}
\newcommand\st{\sin \thetak}
\newcommand\Einf{\mathbf{E}_{\infty}(\thetak,\phik)}

\begin{document}
\title{Helicity and angular momentum. A symmetry based framework for the study of light-matter interactions.}
\author{Ivan Fernandez-Corbaton$^{1,2}$, Xavier Zambrana-Puyalto$^{1,2}$ and Gabriel Molina-Terriza$^{1,2}$}
\affiliation{$^1$ QSciTech and Department of Physics \& Astronomy, Macquarie University, Australia}
\affiliation{$^2$ ARC Center of Excellence for Engineered Quantum Systems}
\begin{abstract}
We propose a new theoretical and practical framework for the study of light-matter interactions and the angular momentum of light. Our proposal is based on helicity, total angular momentum, and the use of symmetries. We compare the new framework to the current treatment, which is based on separately considering spin angular momentum and orbital angular momentum and using the transfer between the two in physical explanations. In our proposal, the fundamental problem of spin and orbital angular momentum separability is avoided, predictions are made based on the symmetries of the systems, and the practical application of the concepts is straightforward. Finally, the framework is used to show that the concept of spin to orbit transfer applied to focusing and scattering is masking two completely different physical phenomena related to the breaking of different fundamental symmetries: transverse translational symmetry in focusing and electromagnetic duality symmetry in scattering.
\end{abstract}
\maketitle

\section{Introduction}
In the last decades, the angular momentum of light has received much attention from very diverse areas of Physics. From experimental astrophysics proposals for the detection of exotic cosmic objects \cite{Tamburini2011} to the use of light beams to rotate atoms \cite{Andersen2006}, through the exploitation of the infinite number of possible angular momentum values for increasing the capacity of optical communication networks \cite{Djordjevic2011} or developing new concepts in quantum information \cite{Gabi2004}. The list of applications is long indeed \cite{Torres2011}.

The availability of appropriate theoretical tools for the study of light beams with angular momentum and their interactions with matter is crucial for the development of such a wide and promising range of applications. The current state-of-the-art theoretical framework is based on the separation of angular momentum ${\mathbf{J}}$ in its orbital (${\mathbf{L}}$) and spin (${\mathbf{S}}$) components: ${\mathbf{J}}={\mathbf{L}}+{\mathbf{S}}$. In the paraxial approximation the value of the total angular momentum along the optical axis can be split into a term which depends on the azimuthal spatial phase of the field and a term which depends on the polarization \cite{Allen1992}.  Several efforts have been undertaken to rigorously extend this approach to the non-paraxial regime, but have encountered fundamental difficulties \cite{Barnett1994,Allen1999}. Strongly non-paraxial tightly focused light fields are the bread and butter of applications where light is made to interact with nano-structures, molecules and atoms. The mechanism of spin to orbit angular momentum conversion (SAM to OAM), also referred to as spin-orbit interaction, is the explanation of choice for numerically obtained observations in focusing \cite{Zhao2007,Nieminen2008,Yao2011} and remarkable results in scattering experiments \cite{Gorodetski2009,Vuong2010,Manni2011}. For example, the presence of optical vortices in tightly focused fields and in scattered fields is explained by conversions between the two types of angular momenta.

Rigorously speaking, though, the separation between SAM and OAM cannot be made on firm physical grounds. Consequently, a conversion between the two quantities is not a fully satisfactory explanation for physical phenomena. The separate consideration of ${\mathbf{S}}$ and ${\mathbf{L}}$ is known to pose fundamental problems for the electromagnetic field \cite{Barnett1994,Allen1999} and its quantum excitations \cite[\S 16]{Berestetskii1982}, \cite[p.50]{Cohen1997}. From the point of view of quantum field theory only the total angular momentum operator is a valid observable property of the photon. Even more generally, the strict non-separability is not restricted to photons. For example, it also applies to the electron since only in the non-relativistic limit can the orbital and spin parts of its angular momentum be separately considered \cite[\S 16]{Berestetskii1982}. The fundamental reason for such non-separability is the geometry of rotations of vectors and spinors. Such a restriction also applies to rotations of classical electromagnetic fields \cite{Rose1957}.

In this article we put forward an alternative theoretical framework for the general and rigorous treatment of the angular momentum of light and its role in light-matter interactions. Our approach solves the theoretical difficulties of the current framework, draws its predictive power from symmetry considerations, and can be simply applied in practice. By using it, we discover that the actual physical reason responsible for the presence of optical vortices in tightly focused beams is totally unrelated to the one responsible for the appearance of optical vortices in scattering experiments. In the current state-of-the-art framework, both cases are explained as SAM to OAM conversion. 

Our proposal is based on total angular momentum and helicity. The role of helicity ($\Lambda$) in light matter interactions has recently been considered \cite{FerCor2012}. The macroscopic Maxwell equations have been shown to be invariant under generalized electromagnetic duality transformations, and helicity has been identified as the generator of those transformations. By exploiting this connection, helicity can be used within the powerful formalism of symmetries and conserved quantities for the study of light matter interactions when the approximations implicit in the macroscopic Maxwell equations hold \cite[chap. 6]{Jackson1998}. The use of symmetries and conserved quantities for the study of electromagnetic problems is the paradigm used in this article for the development of its theoretical concepts and their application to practical situations.

In section \ref{sec:context} we outline the paradigm, mathematical concepts and notation used throughout the paper. In section \ref{sec:LzSz} we summarize the different aspects involved in the separation of SAM and OAM. In section \ref{sec:framework} we outline our proposal. First, we discuss the concept of helicity and its associated symmetry and comment on a result from \cite{FerCor2012}, which shows that, upon scattering, helicity transforms independently of the geometry of the scatterer. Then we show that the combined use of angular momentum and helicity solves the problems associated with the separation of SAM and OAM in a way that is simpler than the existing theoretical solutions and comment on the practical applicability of our ideas. Finally, we establish a relationship between helicity eigenstates and the transverse electric (TE) and transverse magnetic (TM) components of the field. Using this relationship we express the conservation law for helicity as a function of the partial scattering matrices in the TE-TM basis. This expression becomes useful in the practical application of the framework.  In section \ref{sec:spintoorbit} we use the developed ideas to revisit the concept of SAM to OAM conversion in focusing and scattering.  We are able to clearly identify the underlying reasons for the presence of optical vortices in focused and scattered fields, which happen to be two totally different physical phenomena connected to the breaking of two independent fundamental symmetries: transverse translational symmetry in focusing and electromagnetic duality in scattering. Up to now, the two are explained by SAM to OAM conversion. During section \ref{sec:spintoorbit}, we provide the analytical tools necessary for the practical application of our framework. Section \ref{sec:conc} contains our conclusions and discussion.

\section{Paradigm, mathematical setting and notation}\label{sec:context} 
The paradigm that we follow in our work is the use of symmetries and conserved quantities for the study of electromagnetic problems. In this article we consider classical Maxwell fields. Symmetry operations like rotations and translations are linear transformations that apply to the fields. Similarly, we model the light-matter interactions as linear transformations of the free space fields. The fields themselves will hence always be transverse. These ideas are best formalized using the mathematical setting of Hilbert spaces.

Therefore, in this article we make extensive use of the concepts and tools associated with a vector space endowed with an inner product, i.e. a Hilbert space, and the linear operators acting within that vector space. The vector space we consider is the space of transverse solutions of Maxwell's equations, or transverse Maxwell fields, which we call $\mathbb{M}$. A linear operator within $\mathbb{M}$ takes one of its vectors, a transverse Maxwell field, and transforms it into another vector of the space, still a transverse solution of Maxwell's equations. We are interested in symmetry transformations of the fields: time translation, spatial translations and rotations, etc. These transformations are operators acting within $\mathbb{M}$. Such continuous symmetries are generated by hermitian operators, also acting within $\mathbb{M}$, which are associated with properties of the fields. For instance, energy generates time translations, the components of linear momentum generate spatial translations and the components of angular momentum generate spatial rotations. A hermitian operator $O$ generates a continuous symmetry transformation $T(\theta)$ by means of 
\begin{equation}
T(\theta)=\exp(-i\theta O)=\sum_{n} \frac{(-i\theta O)^n}{n!}.
\end{equation}

See \cite[\S.21]{Rose1957} for the detailed study of classical Maxwell fields using angular momentum and its generated transformation, spatial rotations. The fact that $\mathbb{M}$ has an inner product allows us to speak of hermitian operators. It also allows us to construct orthonormal basis into which any transverse Maxwell field can be expanded. The basis vectors can be chosen to be transverse fields which are simultaneous eigenvectors of four commuting hermitian operators. The choice of the inner product is not necessarily unique, even though the one we have used, defined in section \ref{sec:scattering} has a long tradition for vector fields \cite[expr. 13.1.21]{Morse1953}. 

The consideration of $\mathbb{M}$, the hermitian operators associated with the fundamental quantities of the field and the transformations that these operators generate, allows to study electromagnetic problems using Maxwell fields together with the powerful framework of symmetries and conserved quantities. When the electromagnetic equations of a given system are invariant under the transformations generated by a given hermitian operator, the property of the field associated with that operator is a constant of the motion. Conversely, if the system does not possess that symmetry, we know that the associated property can, in general, change during evolution. The effect of the symmetry of the system is even stronger, as it must also preserve the eigenvectors and eigenvalues of the operator generating the symmetry, thus offering an adequate basis of vectors to solve the electromagnetic problem. As we will show, this is a simple yet insightful approach to electromagnetic problems. In this article, we want to exploit this approach for the study of the angular momentum of light. Consequently, for our purposes, we may only use properties of the electromagnetic field which are associated with a hermitian operator in $\mathbb{M}$. Only then can we consider their associated symmetry. This rules out the separate use of the components of ${\mathbf{L}}$ and ${\mathbf{S}}$, since their associated operators transform a transverse Maxwell field into a non-transverse field \cite{VanEnk1994},\cite[B$_I.2$]{Cohen1997}: they do not act within the required vector space. From this point of view, ${\mathbf{L}}$ and ${\mathbf{S}}$ are qualitatively different from ${\mathbf{J}}$, the linear momentum ${\mathbf{P}}$, the energy $H$ or, as we will discuss, helicity $\Lambda$. In the most commonly used representation, the expressions corresponding to these operators are
\begin{align}
\label{eq:line1}
H=i\frac{\partial}{\partial t},\ \mathbf{P}=-i\nabla &,\ \mathbf{J}=\mathbf{L}+\mathbf{S},\ \Lambda=\frac{\mathbf{J}\cdot\mathbf{P}}{|\mathbf{P}|},\\
\label{eq:line2}
\mathbf{L}=-i\mathbf{r}\times \nabla &,\ S^k_{nm}=-i\epsilon_{knm},
\end{align}
where $S^k$, the $k$-th component of $\mathbf{S}$, is a matrix of indexes $nm$ defined using the totally antisymmetric tensor $\epsilon_{knm}$ with $\epsilon_{123}=1$. As discussed above, the operators in equation (\ref{eq:line1}) transform a transverse Maxwell field into a transverse Maxwell field, while those in equation (\ref{eq:line2}) break the transversality of the fields.

In our notation we use capital letters like $J_z$ and $P_x$ to denote operators, and lower case letters like $j_z$ (or $n$) and $k_x$ to denote their eigenvalues. When we speak of a field having a ``sharp'' or ``well defined'' value for an operator, we mean that the field is an eigenvector of that operator with eigenvalue equal to its ``sharp'' value. In the text, names like ``helicity'' or ``third component of angular momentum'' refer to the operators unless it is clear from the context that this is not the case. Also, all analytical calculations from section \ref{sec:scattering} on, assume a time harmonic decomposition of the fields with an $\exp(-iwt)$ dependence. 

Finally, we would like to mention that the main context of this work is that of classical Maxwell fields. Nevertheless, the approach here taken, which is based on the study of symmetries, is general and often used in Quantum Mechanics and other areas of modern Physics.

\section{The separation of $\mathbf{L}$ and $\mathbf{S}$ }\label{sec:LzSz}
Serious theoretical difficulties are encountered when attempting to separately consider ${\mathbf{L}}$ and ${\mathbf{S}}$ for the electromagnetic fields (\cite{Barnett1994}, \cite{Allen1999}) or its quanta (\cite[\S16]{Berestetskii1982},\cite[B$_I.2$]{Cohen1997}). We also know \cite{VanEnk1994}, \cite[B$_I.2$]{Cohen1997} that the operators $\mathbf{L}$ and $\mathbf{S}$ break the transversality of the fields, taking a vector of $\mathbb{M}$ out of that space. As far as we are concerned, this prevents the separate use of $\mathbf{L}$ and $\mathbf{S}$ for studying electromagnetic problems using symmetries and conserved quantities: $\mathbf{L}$ and $\mathbf{S}$ do not generate symmetry transformations for vectors in $\mathbb{M}$.

On the other hand, the total integrated value of the angular momentum of the electromagnetic field, expressed here in convenient units as an integral over all space involving the electric and magnetic fields $\mathbf{E}(\mathbf{r})$ and $\mathbf{B}(\mathbf{r})$

\begin{equation}
\label{eq:jmean}
\langle {\mathbf{J}} \rangle=\int d\mathbf{r}\ \mathbf{r}\times (\mathbf{E}(\mathbf{r})\times\mathbf{B}(\mathbf{r})),
\end{equation}
can be separated into two gauge invariant integrals. To achieve such separation one needs to consider only the transverse parts of the field, which are always gauge independent. This restriction is justifiable because the degrees of freedom associated with the longitudinal part of the electric field can always be combined with the degrees of freedom of the sources \cite[I.B.5]{Cohen1997},\cite[chap. XXI,\S22]{Messiah1958}. Then, in

\begin{align}
\label{eq:jtrans}
\langle {\mathbf{J_t}} \rangle&=\int d\mathbf{r}\ \mathbf{r}\times (\mathbf{E}_t(\mathbf{r})\times\mathbf{B}(\mathbf{r}))\\
\label{eq:OAMSAM}
&=\int d\mathbf{r}\ (\mathbf{E}_t(\mathbf{r})\times\mathbf{A}_t(\mathbf{r}))+\int d\mathbf{r}\ \sum_{i=1}^3 E_t^i(\mathbf{r})(\mathbf{r}\times\nabla) A^i_t(\mathbf{r})=\langle {\mathbf{S_t}} \rangle + \langle {\mathbf{L_t}} \rangle.
\end{align}
where index $i$ refers to the three spatial components, the two terms $\langle {\mathbf{S_t}}\rangle $ and  $\langle {\mathbf{L_t}} \rangle$ are apparently gauge invariant. From now on, we will drop the $\mathbf{_t}$ underscripts since in this article we only consider transverse Maxwell fields. The identification of the two parts of equation (\ref{eq:OAMSAM}) with spin and orbital angular momenta is tempting due to the appearance of the operator $\mathbf{L}=-i\mathbf{r}\times\nabla$ and the relationship of the cross-product in $\langle {\mathbf{S_t}} \rangle$ with the spin-1 matrices representing $\mathbf{S}$. But, since $\mathbf{L}$ and $\mathbf{S}$ are not operators in $\mathbb{M}$, the question arises of which are the operators corresponding to the two parts of equation (\ref{eq:OAMSAM}). In his book on Quantum Mechanics, Messiah \cite[Ch. XXI, problem 7]{Messiah1958} offers an expression which corresponds to the second quantization of the first part of equation (\ref{eq:OAMSAM}). In 1994, Van Enk and Nienhuis \cite{VanEnk1994}, in a more detailed study, arrived at the same result and also derived the expression for the second part of equation (\ref{eq:OAMSAM}). In that work, they showed that the two operators are not angular momenta because they do not satisfy the commutation relations which define angular momentum operators. These Fock space operators have their corresponding operators in $\mathbb{M}$ for classical fields (see section \ref{sec:theorprac}), and their third components commute \cite{Bliokh2010}. It is hence possible to split the total angular momentum $\mathbf{J}$ into two operators $\mathbf{\widehat{S}}=\Lambda \frac{\mathbf{P}}{|\mathbf{P}|}$ and $\mathbf{\widehat{L}}=\mathbf{J}-\mathbf{\widehat{S}}$. Consequently $\langle\mathbf{L}_t\rangle$ and $\langle\mathbf{S}_t\rangle$ in (\ref{eq:OAMSAM}) are in reality $\langle\mathbf{\widehat{L}}\rangle$ and $\langle\mathbf{\widehat{S}}\rangle$. Unfortunately, in terms of the separation of the angular momentum operator, this approach is not fully satisfactory because the resulting operators are not angular momentum operators. This can be easily proved by checking that their vectorial components do not fulfill the commutation relations required for angular momentum operators. The consequence of this is that they cannot separately generate meaningful rotations. In summary, $\mathbf{J}$ can always be decomposed into two meaningful operators, $\mathbf{J}=\mathbf{\widehat{L}}+\mathbf{\widehat{S}}$, independently of the paraxial approximation. These two operators are never angular momenta since they never fulfill the correct commutation relationships. This last statement is also independent of the paraxial approximation. See section \ref{sec:theorprac} for a more detailed discussion.

Following the paradigm discussed in section \ref{sec:context}, we only consider properties of the field whose corresponding operators generate transformations in $\mathbb{M}$. As already mentioned several times, this disqualifies $\mathbf{L}$ and $\mathbf{S}$. As substitutes, one may choose $\mathbf{\widehat{L}}$ and $\mathbf{\widehat{S}}$. We prefer to disregard the question of the separation completely and use the total angular momentum $\mathbf{J}$ and the helicity instead. Our choice is based on the fact shown below (section \ref{sec:theorprac}) that $\mathbf{J}$ and helicity generate very simple symmetry transformations which lead to a simpler framework.


Finally, it is worth noting that expressions (\ref{eq:jmean}) and (\ref{eq:jtrans}) can be interpreted and computed as a weighted average. When the electromagnetic field is decomposed in a basis of eigenvectors of $J_z$, the total integrated value $\langle J_z \rangle$ is equal to a weighted average \cite[chap. 9.8, expr. 9.143]{Jackson1998}. The weights are the squares of the amplitudes in the linear decomposition of the field, and they multiply the different eigenvalues $j_z$ of each mode in the basis. This connection is not restricted to angular momentum. All the integrals that are used to compute total integrated values of properties of the electromagnetic field can be interpreted and computed as weighted averages involving the squares of the expansion coefficients and the eigenvalues of the operator related to each particular property. This connection is the key step used for obtaining a Fock space operator from the classical spatial integral involving the fields. See for instance \cite[chap. 10.2.3, 10.5]{Mandel1995} for the cases of  $H$ and $\mathbf{P}$ respectively. This connection relates the algebraic formalism introduced in sec. \ref{sec:context} with the well known spatial integrals involving the fields, like (\ref{eq:jmean}) for the case of angular momentum.



\section{A framework based on helicity and angular momentum}\label{sec:framework}
\subsection{The helicity of light fields and its associated symmetry}
\begin{figure}[ht]
\begin{center}
\includegraphics{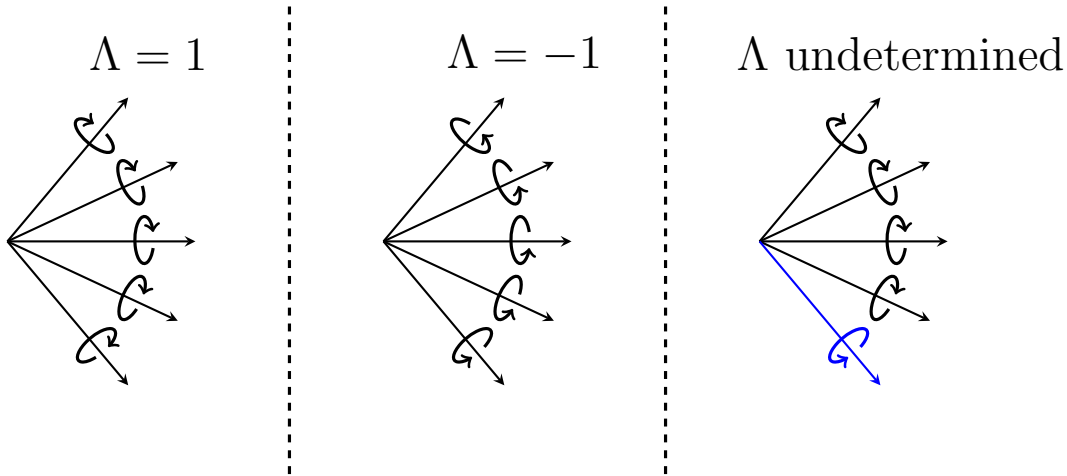}
\end{center}
\caption{(Color online) A field composed by the superposition of five plane waves has definite helicity equal to one if all the plane waves have left handedness (left part), equal to minus one if they all have right handedness (central part) and does not have a definite helicity if all the plane waves do not have the same handedness (right part).}
\label{fig:helicity}
\end{figure}

Helicity is defined as the operator which results from projecting the total angular momentum onto the linear momentum \cite[chap. 8.4.1]{Tung1985}, i.e. $\Lambda={\mathbf{J}}\cdot{\mathbf{P}}/|{\mathbf{P}}|$. Helicity commutes with all the generators of rotations ${\mathbf{J}}$ and translations ${\mathbf{P}}$ \cite[chap. 10.4.3]{Tung1985}. In the case of the photon, the helicity can only take the values $\pm1$ \cite[chap. 2.5]{Weinberg1995}. For the electromagnetic field, $\Lambda$ has only two eigenvalues equal to $\pm 1$. A useful interpretation of helicity is obtained by considering the expansion of the field as a superposition of plane waves. In such a representation, helicity is associated with the handedness of each plane wave. For a particle to have a well defined helicity, all the plane waves must be purely circularly polarized and have the same handedness with respect to its momentum vector. This is illustrated in Fig. \ref{fig:helicity}.

Given that a general light beam can always be expanded as a sum of plane waves and that helicity is associated with the polarization handedness of those plane waves, helicity seems a very suitable candidate for representing the polarization degrees of freedom of a general light beam. Crucially, the action of helicity on a plane wave affects only its polarization vector: it leaves the plane wave momentum vector invariant. This is also true for other types of light beams like multipolar fields or Bessel beams: the action of helicity does not change the quantities related to energy, linear or angular momenta which define those beams. Helicity acts on separated degrees of freedom, which, in this article, we refer to as the degrees of freedom of the polarization.

From its definition, it is readily checked that helicity is a hermitian operator. With respect to related transformations, helicity is the generator of the generalized electromagnetic duality transformation of the fields:
\begin{equation}
\label{eq:DualityGeneralization}
\begin{split}
\mathbf{E}&\rightarrow \cos(\theta) \mathbf{E} - \sin(\theta)\mathbf{H},\\
\mathbf{H}&\rightarrow \sin(\theta) \mathbf{E} + \cos(\theta)\mathbf{H},
\end{split}
\end{equation}
where $\theta$ is a real angle. That helicity generates duality is a remarkable fact which can hardly be deduced from the mathematical definition of helicity containing the angular and linear momenta.

Generalized duality (\ref{eq:DualityGeneralization}) is well known to be a symmetry of the source free microscopic Maxwell equations. The results in \cite{FerCor2012} allow to extend the relationship between helicity and duality to the macroscopic Maxwell equations in material systems. That work shows that the helicity of the light interacting with a piecewise homogeneous isotropic medium will transform independently of the shapes of the different material domains, and that helicity will be preserved by the interaction if and only if the ratio between the relative electric and magnetic constants of all the involved materials $i\in[1\ldots N]$ is constant, that is:
\begin{equation}
\label{eq:epsmu}
\frac{\epsilon_i}{\mu_i}=\alpha\ \forall \ i.
\end{equation}

When condition (\ref{eq:epsmu}) is met, the macroscopic Maxwell's equations for the system are invariant under transformation (\ref{eq:DualityGeneralization}). 

The geometry independent character of the duality symmetry (helicity conservation) allows to separately consider the transformations of the polarization degrees of freedom from the geometry of the scattering problem. Such notable simplification is very desirable in a framework for the study of light matter interactions.

We propose the use of helicity, the generator of duality transformations, for treating the polarization degrees of freedom in electromagnetic problems. We will show that, when used in conjunction with the total angular momentum, the generator of rotations, we obtain a general framework for the study of problems involving the angular momentum of light and its interaction with matter using the language of symmetries and conserved quantities. It is important to recall that, in this article, we will remain within the approximations implicit in the macroscopic Maxwell's equations \cite[chap. 6]{Jackson1998}.

For completeness, a expression of the total integrated value of helicity in terms of electromagnetic fields can be found in \cite{Afanasiev1996}. In this article, though, we are not concerned with the total integrated value of helicity. Instead, we use modes with well defined helicity and consider their transformations in particular situations, which will depend on whether the system is invariant under electromagnetic duality.

\subsection{Theoretical and practical considerations when using $\Lambda$ and $J_z$}\label{sec:theorprac}

The fundamental problem of the separability of $S_z$ and $L_z$ is avoided by using $\Lambda$ and $J_z$ instead. Both $J_z$ and $\Lambda$ are commuting hermitian operators in $\mathbb{M}$, and generate two simple and independent symmetry transformations of the electromagnetic field: $J_z$ generates rotations around the $z$ axis and $\Lambda$ generates the generalized duality transformation (\ref{eq:DualityGeneralization}).  

As previously mentioned, the two Fock space operators obtained from the second quantization of expression (\ref{eq:OAMSAM}) studied in \cite{VanEnk1994} (and also later in \cite{Jauregui2005}) have their corresponding operators for classical fields: in \cite{Bliokh2010} they are obtained in the momentum space. For instance, it can be seen that $\widehat{S}_z$, the operator substituting $S_z$ is $\Lambda P_z/|\mathbf{P}|$. This operator commutes with $\widehat{L}_z$ \cite{Jauregui2005,Bliokh2010}, whose expression is obviously $\widehat{L}_z=J_z-\Lambda P_z/|\mathbf{P}|$. Consequently, we could use $\widehat{S}_z$ and $\widehat{L}_z$ instead of $J_z$ and $\Lambda$. Considering the symmetries generated by each pair of operators we prefer to choose $J_z$ and $\Lambda$ for reasons of simplicity. While $J_z$ and $\Lambda$ are related to the two simple symmetries indicated above, the symmetries related with $\widehat{S}_z$ and $\widehat{L}_z$ are more complicated. An explicit mathematical expression for the transformation generated by $\mathbf{\widehat{S}}$ can be found in \cite{Cameron2012}. For once, $\widehat{S}_z$ involves a combination of duality and translational symmetries, while the symmetry related to $\widehat{L}_z$ has not been properly studied as such, up to our knowledge. This suggests that using $J_z$ and $\Lambda$ should, in most situations, result in a simpler analysis. For instance, it is explicitly seen in section \ref{sec:focusing} that the conservation law associated with $\Lambda P_z/|\mathbf{P}|$ is not independent of the geometry of the problem.

In the paraxial limit, when $P_z\rightarrow P$, it can be shown that $\widehat{S}_z\rightarrow \Lambda$ and $\widehat{L}_z\rightarrow J_z-\Lambda$. Even in this limit, neither of these operators generates physical rotations for the full electromagnetic field. For a paraxial beam, the helicity may be approximated by the real space circular polarization component perpendicular to the $z$ axis and, the value of $J_z-\Lambda$ coincides (see equations (\ref{eq:Cparaxial})-(\ref{eq:Dparaxial}) below) with the azimuthal phase winding number of the dominant circular polarization. It is customary to use the paraxial correspondence between SAM and circular polarization and OAM and azimuthal phase. From the previous considerations, we think that it would be more insightful to use $\Lambda$ and $J_z-\Lambda$, which have the advantage of retaining its meaning outside the paraxial approximation.

From the experimental point of view, section \ref{sec:helvor} shows that the preparation of general (non-paraxial) beams with well defined helicity can be done using simple optics in a straightforward fashion. The measurement of the helicity state of a general beam is also shown to be easily achieved using simple optics.

In this article we focus in the combined use of $\Lambda$ and $J_z$. However, it is worth mentioning that the fact that $\Lambda$ commutes with all the generators of rotations ${\mathbf{J}}$ and translations ${\mathbf{P}}$, and that it generates a transformation which is independent of geometry, should allow its combined use with other degrees of freedom different from angular momentum when the particular problem requires it.

\subsection{Helicity preservation conditions on the TE-TM scattering matrices}\label{sec:scattering}
In this section we arrive at a relationship between helicity conservation and the scattering of the transverse electric and transverse magnetic components of the field. This relationship is useful in practical problems, as will be seen in section \ref{sec:spintoorbit}.

\begin{figure}[htbp]
\centering\includegraphics[width=7cm]{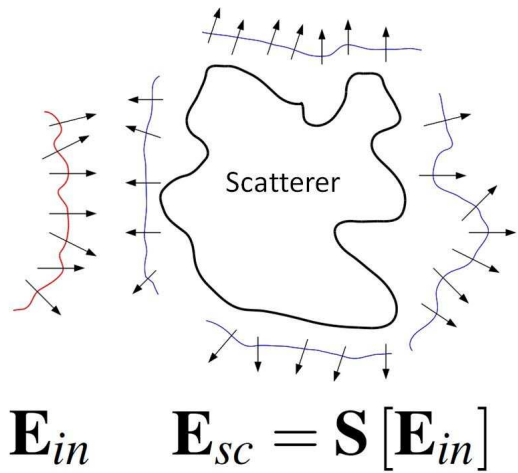}
\caption{(Color online) Scattering problem of arbitrary geometry}
\label{fig_scatt}
\end{figure}

Let us consider the general scattering problem presented in figure (\ref{fig_scatt}). An incident electromagnetic field $\mathbf{E}_{in}$ impinges onto a scatterer of arbitrary shape $S$ resulting in the scattered field $\mathbf{E}_{sc}$. Consider an orthonormal basis of transverse electromagnetic modes with well defined helicity $\lbrace \Aplus,\Aminus \rbrace$ $\forall$ $\nu$, where the superindex denotes the sign of $\Lambda$ and the subindex $\nu$ is a composed index which contains three other commuting operators. For example $\nu=[H,P_x,P_y]$ for plane waves, with $H$ being the energy, and $P_x,\ P_y$ the first two components of linear momentum.

Both incident and scattered fields can be expanded using the $\lbrace \Aplus,\Aminus \rbrace$ basis, and the scattering process is fully characterized by the following transformation of the incident field:
\begin{equation}
\label{eq:scatt}
\begin{split}
\mathbf{E}_{sc}&=\mathbf{S}\mathbf{E}_{in},\\
\mathbf{E}_{sc}&=\left[\int d\bar{\nu}\int d\nu S_{\nu,+}^{\bar{\nu},+} \Aplusbar\Aplusdagger + S_{\nu,+}^{\bar{\nu},-} \Aminusbar\Aplusdagger + S_{\nu,-}^{\bar{\nu},+} \Aplusbar\Aminusdagger + S_{\nu,-}^{\bar{\nu},-} \Aminusbar\Aminusdagger\right]\mathbf{E}_{in},
\end{split}
\end{equation}
where ${\mathbf{A}}^\dagger$ is the hermitian conjugate of $\mathbf{A}$. In (\ref{eq:scatt}), the transformation $\mathbf{S}$ is represented by a doubly infinite weighted sum of projection operators of the type ${\mathbf{A}}{\mathbf{A}}^\dagger$, whose action on the input field is $({\mathbf{A}}{\mathbf{A}}^\dagger) \mathbf{E}_{sc}={\mathbf{A}}({\mathbf{A}}^\dagger\mathbf{E}_{in})$ and the inner product
\begin{equation}
\label{eq:inner}
{\mathbf{A}}^\dagger\mathbf{E}_{in}=\int d\mathbf{r} {\mathbf{A}(\mathbf{r})}^\dagger \mathbf{E}_{in}(\mathbf{r}),
\end{equation}
is assumed.

Transformation (\ref{eq:scatt}) is specified by the infinite set of $2\times 2$ partial scattering matrices relating the coefficients of $\mathbf{E}_{in}$ in $\Aplus,\Aminus$ to the coefficients of $\mathbf{E}_{sc}$ in $\Aplusbar,\Aminusbar$ for all $\nu$, $\bar{\nu}$:

\begin{equation}
\label{eq:scatthel}
S_\nu^{\bar{\nu}}=
\begin{pmatrix}
S_{\nu,+}^{\bar{\nu},+} & S_{\nu,-}^{\bar{\nu},+}\\ S_{\nu,+}^{\bar{\nu},-}& S_{\nu,-}^{\bar{\nu},-}\\
\end{pmatrix}.
\end{equation}

Let us impose helicity conservation upon scattering, forcing all the partial scattering matrices to be diagonal: $S_{\nu,-}^{\bar{\nu},+}= S_{\nu,+}^{\bar{\nu},-}=0$ for all $\nu$, $\bar{\nu}$. 

Consider the following linear combinations of modes with well defined helicity.
\begin{equation}
\label{eq:MNAA}
\Mreta = \frac{1}{\sqrt{2}}\left(\Aplus + \Aminus\right),\ \Nreta = \frac{1}{\sqrt{2}}\left(\Aplus - \Aminus\right),
\end{equation}
which transform into each other by the action of $\Lambda$
\begin{equation}
\Lambda \Mreta = \Nreta,\ \Lambda \Nreta = \Mreta.
\end{equation}

In appendix \ref{appA} we show that, for plane waves, Bessel beams and multipoles, the sum and subtraction of modes with same $\nu$ and different helicity result in what are commonly known as TE and TM modes. We will adopt the TE-TM naming to denote general sum and subtractions of modes differing only by their sharp helicity eigenvalue.

Since $\lbrace \Aplus,\Aminus \rbrace$ is an orthonormal basis, so is $\lbrace \Mreta,\Nreta \rbrace$. After changing (\ref{eq:scatt}) to the $\lbrace \Mreta,\Nreta \rbrace$ basis, the condition for helicity preservation upon scattering, expressed in the TE-TM basis reads:

\begin{equation}
\label{eq:tetmhel}
\begin{split}
\begin{pmatrix}
S_{\nu,TE}^{\bar{\nu},TE} & S_{\nu,TE}^{\bar{\nu},TM}\\ S_{\nu,TM}^{\bar{\nu},TE}& S_{\nu,TM}^{\bar{\nu},TM}
\end{pmatrix}
&=
\frac{1}{2}
\begin{pmatrix}
1 & 1\\1 & -1
\end{pmatrix}
\begin{pmatrix}
S_{\nu,+}^{\bar{\nu},+} & 0\\ 0 & S_{\nu,-}^{\bar{\nu},-}\\
\end{pmatrix}
{\begin{pmatrix}
1 & 1\\1 & -1
\end{pmatrix}}^{-1}=\\
&=\frac{1}{2}
\begin{pmatrix}
S_{\nu,+}^{\bar{\nu},+}+S_{\nu,-}^{\bar{\nu},-} & S_{\nu,+}^{\bar{\nu},+}-S_{\nu,-}^{\bar{\nu},-}\\
S_{\nu,+}^{\bar{\nu},+}-S_{\nu,-}^{\bar{\nu},-} & S_{\nu,+}^{\bar{\nu},+}+S_{\nu,-}^{\bar{\nu},-}\\
\end{pmatrix}
=
\begin{pmatrix}
a_{\nu}^{\bar{\nu}} & b_{\nu}^{\bar{\nu}}\\ b_{\nu}^{\bar{\nu}}& a_{\nu}^{\bar{\nu}}
\end{pmatrix},
\end{split}
\end{equation}
for all ($\nu$, $\bar{\nu}$). Namely, that the scattering of TE and TM components is on an equal footing for all ($\nu$, $\bar{\nu}$), as can be seen from the scattering matrix having the same values $a_{\nu}^{\bar{\nu}}$ in the diagonal and $b_{\nu}^{\bar{\nu}}$ in the off-diagonal. Condition (\ref{eq:tetmhel}) is clearly a restriction which will not be met in general. We conclude that, in general, a scatterer will partially convert the helicity of the incident field and that this partial helicity conversion is reflected in asymmetries of the scattering matrices with respect to the TE and TM modes. 

Since helicity conservation is equivalent to invariance under generalized duality transformations, having information on the TE-TM scattering properties of a system can be used to assess its duality invariance and vice versa.

In systems with a high degree of symmetry, like a sphere or a planar multilayer system, a wise choice of basis simplifies condition (\ref{eq:tetmhel}). The symmetries of those two systems make the TE and TM multipoles \cite[chap. 9]{Jackson1998} and the TE and TM plane waves (definition contained in appendix \ref{appB}) the eigenmodes of the spherical and planar structures respectively. This means that $S_\nu^{\bar{\nu}}=0$ unless $\nu=\bar{\nu}$ and that $S_{\nu,TE}^{{\nu},TM}=S_{\nu,TM}^{{\nu},TE}=0$ for all $\nu$, $\bar{\nu}$. All the partial scattering matrices are diagonal. The preservation of the TE-TM components is related to the geometrical symmetries of the system. Using Mie scattering theory and Fresnel's formulas, it is an easy and interesting exercise to analytically verify that, in these two cases, when the materials meet (\ref{eq:epsmu}), all the scattering matrices are proportional to the identity ($S_{\nu,TE}^{{\nu},TE}=S_{\nu,TM}^{{\nu},TM}$ for all $\nu$), hence preserving helicity as well. 

\section{Exemplary application of the framework: revision of spin to orbit angular momentum conversion in focusing and scattering}\label{sec:spintoorbit}
The conversion between spin and orbital angular momentum is widely used to explain phase singularities in numerical simulations of tightly focused fields \cite{Zhao2007,Nieminen2008,Yao2011}, and in scattering experiments: \cite{Gorodetski2009,Vuong2010,Manni2011}. A detailed discussion of the SAM to OAM conversion can be found in \cite{Bliokh2011}.

We will now use symmetries and conserved quantities arguments, particularly those related to $J_z$ and $\Lambda$, to identify the actual physical reasons for those observations. We will demonstrate that the mechanism responsible for the presence of optical vortices in focused fields is totally different from the one responsible for the observation of optical vortices in scattering experiments. This, in our opinion, shows that the SAM-OAM formulation can be quite misleading: it assigns the same explanation to two distinct physical phenomena.

The analytical tools and methodology employed in this section allow a simple application of the ideas developed in the previous section to practical electromagnetic problems. 

\subsection{Bessel beams with well defined angular momentum and helicity}
As already mentioned, $J_z$ and $\Lambda$ commute. For our analysis we will need a basis of transverse electromagnetic modes which are simultaneous eigenvectors of $J_z$ and $\Lambda$. One such set of modes is a particular type of Bessel beams. Bessel beams have been thoroughly studied. See for instance the series of papers \cite{Jauregui2005}, \cite{Jauregui2004}, and \cite{Hacyan2006}. 

In appendix \ref{appB}, we constructively derive a complete orthonormal basis for transverse electromagnetic fields consisting of Bessel beams with well defined third component of angular momentum $J_z$ and helicity $\Lambda$. These modes appear in \cite{VanEnk1994}, although their relationship with helicity is not considered in that paper. In \cite{Jauregui2005}, these type of electromagnetic modes are obtained in a different way as linear combinations of other type of more commonly used Bessel beams, the transverse electric (TE) and transverse magnetic (TM) modes. The constructive derivation in appendix \ref{appB} shows clearly that the modes are eigenstates of $\Lambda$. 

From appendix \ref{appB}, equations (\ref{eq:CD}) are the real space expressions in cylindrical coordinates $[\rho,\theta,z]$. An implicitly harmonic $\exp(-i w t)$ dependence has been assumed.

{\scriptsize 
\begin{equation}
\begin{split}
\label{eq:CD}
\Cnkrho(\rho,\theta,z)&=\sqrt{\frac{\krho}{2\pi}}i^n\exp(i(\kz z+n\theta))\left(\frac{i}{\sqrt{2}}\left((1+\frac{\kz}{k})J_{n+1}(\krho\rho)\exp(i\theta)\rhat+(1-\frac{\kz}{k})J_{n-1}(\krho\rho)\exp(-i\theta)\lhat \right)-\frac{\krho}{k}J_n(\krho\rho)\zhat\right),\\
\Dnkrho(\rho,\theta,z)&=\sqrt{\frac{\krho}{2\pi}}i^n\exp(i(\kz z+n\theta))\left(\frac{i}{\sqrt{2}}\left((1-\frac{\kz}{k})J_{n+1}(\krho\rho)\exp(i\theta)\rhat+(1+\frac{\kz}{k})J_{n-1}(\krho\rho)\exp(-i\theta)\lhat \right)+\frac{\krho}{k}J_n(\krho\rho)\zhat\right),\\
\end{split}
\end{equation}
}
where:
\begin{itemize}
\item $\krho=\sqrt{k_x^2+k_y^2}$, $k^2=\krho^2+\kz^2$
\item $\lhat=\frac{\xhat + i \yhat}{\sqrt{2}}$, $\rhat=\frac{\xhat - i \yhat}{\sqrt{2}}$.
\end{itemize}

By construction, the two types of vector wave functions $\Cnkrho$ and $\Dnkrho$ have a sharp value of the $z$ component of angular momentum $J_z$ equal to $n$ and a sharp value of helicity $\Lambda$ equal to $-1$ and $+1$, respectively. Additionally, they have well defined values for the energy $H$ and the $z$ component of the linear momentum $P_z$ proportional to $k$ and $\kz$ respectively. For a given value of $k$, a well defined value of $P_z$ implies a well defined value of the transverse momentum $P_\rho$ proportional to $\krho$.  
Modes (\ref{eq:CD}) form a complete orthonormal basis of transverse electromagnetic fields when
\begin{equation}
k \in [0,\ \infty),\ \ n\in [0,\pm 1,\pm 2,\ldots],\ \ \krho \in [0, \ \infty),\text{and } \Lambda = \pm 1,
\end{equation}
and both signs of $\kz$ in $\kz=\pm\sqrt{k^2-\krho^2}$ are considered. In the following, the consideration of both signs of $\kz$ is implicitly made.

\subsection{Optical vortices in focusing}\label{sec:focusing}

In order to study why optical vortices seem to appear in numerical studies of focalization of apparently vortex free beams \cite{Zhao2007,Nieminen2008,Yao2011}, we analyze the aplanatic lens model \cite{Richards1959}, which is routinely used to study the effects of high numerical aperture lenses.

As we show in appendix \ref{appC} and has been discussed before \cite{Bliokh2011}, the action of an aplanatic lens preserves $J_z$ and $\Lambda$. The cylindrical symmetry of the model can be reasonably expected, but its invariance under duality transformations is ``hidden'' in the assumption that the lens transmission coefficients for the two polarization components, TE and TM, are identical and that there is no crosstalk between input and output TE and TM components. That this assumption implies duality symmetry (helicity conservation) is obvious from condition (\ref{eq:tetmhel}) and the discussion at the end of section \ref{sec:scattering}. The preservation of $\Lambda$ by an aplanatic lens has been discussed in \cite{Bliokh2011} without using its relationship to electromagnetic duality. In appendix \ref{appC} we explicitly analyze the conservation of $J_z$, highlight the model's key assumption on TE and TM scattering, and show that the model conditions for the conservation of $\Lambda$ and $J_z$ are, as expected, independent of each other. 

Even though both cylindrical and generalized duality symmetries are preserved by the model, the focalized beam is quite different from the input beam: some other symmetry must be broken. The most obvious candidate is the lack of translational symmetry on the plane perpendicular to the optical axis of the lens. We know that the transverse momentum components $P_x$ and $P_y$ are the generators of that symmetry. Below we show that the differences between the input and focalized beams are due to changes in $(k_x,k_y)$. We will prove this point using the basis introduced in (\ref{eq:CD}).

Let us take a collimated right circularly (RC) polarized Gaussian beam going through an aplanatic lens of high numerical aperture. The linear momentum components of a collimated beam are all almost totally aligned with the propagation direction $z$: $\kz\approx k$. Consequently, a collimated beam only has components with small transverse momentum value $\krho$. With respect to (\ref{eq:CD}), in the limit when $\kz\approx k$ ($\smallkrho$), both $\Cnkrho$ and $\Dnkrho$ approach pure RC and LC polarized modes respectively.  This can be easily seen by setting a $\smallkrho \Rightarrow (\kz \approx k)$ in (\ref{eq:CD}): 

\begin{align}
\label{eq:Cparaxial}
\Cnkrho(\rho,\theta,z)&\approx\sqrt{\frac{\krho}{\pi}}i^{n+1}\exp(i(\kz z))J_{n+1}(\krho\rho)\exp(i\theta (n+1))\rhat,\\
\label{eq:Dparaxial}
\Dnkrho(\rho,\theta,z)&\approx\sqrt{\frac{\krho}{\pi}}i^{n+1}\exp(i(\kz z))J_{n-1}(\krho\rho)\exp(i\theta (n-1))\lhat.
\end{align}
The other polarization components, the opposite circular and the longitudinal $\zhat$ component, are strongly attenuated in this regime. Importantly though, they are actually present: without them, the modes are not solutions of Maxwell's equations, and its transformation properties cannot be consistently analyzed as general electromagnetic fields.

From (\ref{eq:Cparaxial}), and since the collimated input RC Gaussian beam does not have a phase singularity in its $\rhat$ dominant polarization, we can conclude that mostly $C$ type modes with $n=-1$ will exist in its expansion in the (\ref{eq:CD}) basis:

\begin{equation}
\label{eq:last}
\mathbf{E}_{input}=\intkrho c^{input}_{-1,\krho}\Cnkrho,
\end{equation}
where $c^{input}_{-1,\krho}$ is only significant when $\smallkrho$. Equation (\ref{eq:last}) defines a beam with a sharp value of $J_z$, $n=-1$, and a sharp value of $\Lambda$, $\lambda=-1$. As per the above discussion regarding symmetries, the output beam must also have sharp $J_z$ and $\Lambda$ values of $n=-1$ and $\lambda=-1$. Focusing can hence only change the relative weight of $\krho$ components. Intuitively, modes with higher transverse momentum are needed to expand the field after focusing. 

\begin{equation}
\mathbf {E}_{foc}=\intkrho c^{foc}_{-1,\krho}\Cnkrho,
\end{equation}
which is in line the non-preservation of $(k_x,k_y)$ due to broken transverse translational symmetry. The fact that the change is limited to $\krho=\sqrt{k_x^2+k_y^2}$ could have been foreseen: it stems from the cylindrical symmetry of the model.

Now, let's go back to equation (\ref{eq:CD}) and check the spatial shape of $\Cnkrho$ modes when $n=-1$ and the condition $\smallkrho$ is not met:
\begin{equation}
\begin{split}
\label{eq:C}
\Cnkrho(\rho,\theta,z)=&-\sqrt{\frac{\krho}{2\pi}}i\exp(i\kz z)(\frac{i}{\sqrt{2}}(1+\frac{\kz}{k})J_{0}(\krho\rho)\rhat\\&+\frac{i}{\sqrt{2}}(1-\frac{\kz}{k})J_{-2}(\krho\rho)\exp(-i2\theta)\lhat -\frac{\krho}{k}J_{-1}(\krho\rho)\exp(-i\theta)\zhat),\\
\end{split}
\end{equation}

\begin{figure}[htbp]
\begin{center}
  \subfigure[$\frac{\krho}{k}=0.1$]{%
            \label{figsmall}
			\includegraphics[scale=0.8]{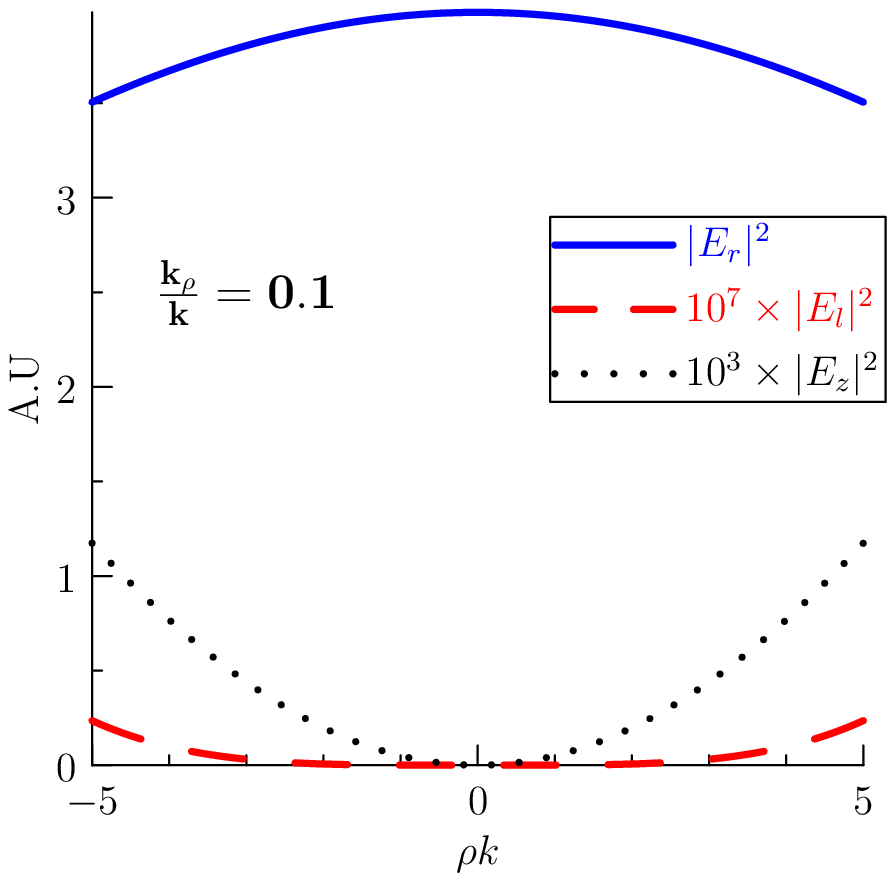}
			}
  \subfigure[$\frac{\krho}{k}=0.9$]{%
            \label{figbig}
			\includegraphics[scale=0.8]{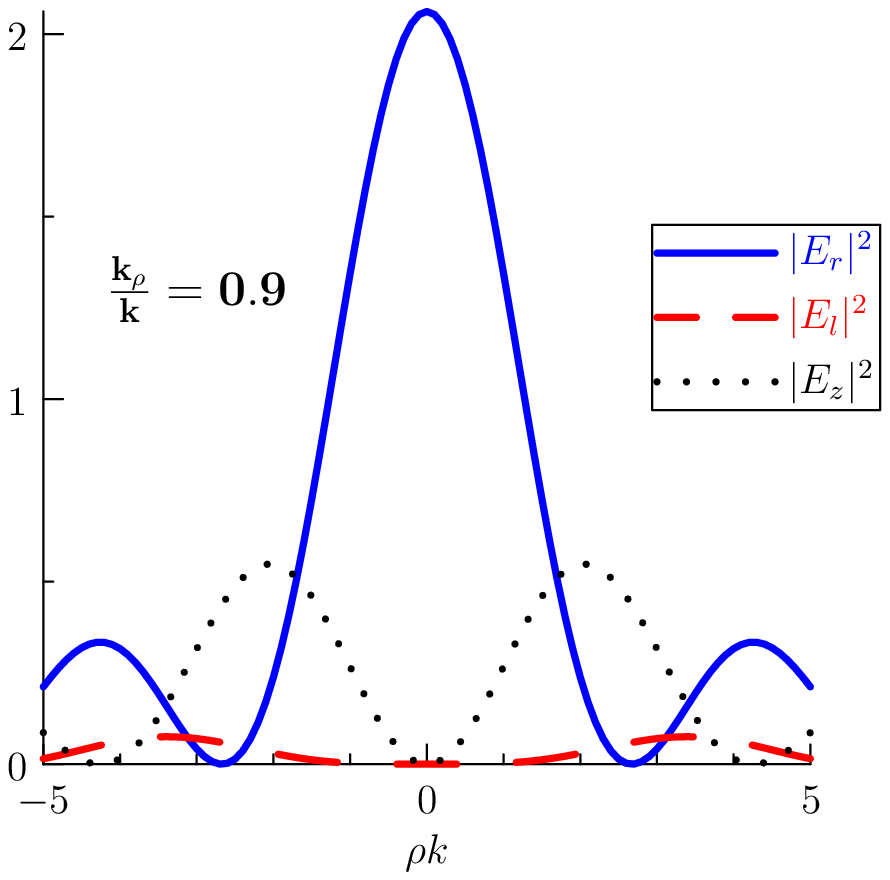}
			}
  \end{center}
\caption{(Color online) Normalized field intensity for the right, left and longitudinal polarization components for two $\Cnkrho$, $n=-1$ modes, one with $\frac{\krho}{k}=0.1$ (left figure) and the other with $\frac{\krho}{k}=0.9$. Note the scaling of the non-dominant polarization components on the $\frac{\krho}{k}=0.1$ case.}
\label{fig:smallbig}
\end{figure}

The $\Cnkrho,\smallkrho$ modes are almost purely right polarized, but when $\krho$ increases, the terms multiplying $\lhat$ and $\zhat$ become significant. As it can be seen in (\ref{eq:C}), for $n=-1$ these newly enhanced terms have phase singularities of orders minus two and minus one respectively. Fig. \ref{fig:smallbig} shows the radial spatial profiles of the three polarization components for two $\mathbf{C}_{-1,\krho}(\rho,\theta,z)$ modes, one with $\frac{\krho}{k}=0.1$ and the other with $\frac{\krho}{k}=0.9$. In the small $\krho$ case (Fig. \ref{figsmall}), the dominant polarization component $\rhat$ is much stronger than the $\lhat$ and $\zhat$ components (which are nonetheless present). In the large $\krho$ case (Fig. \ref{figbig}), the relative weight between the intensity of the different polarizations has shifted significantly. The vortices of charge $-1$ in $\zhat$ and charge $-2$ in $\lhat$ (see (\ref{eq:C})) become relatively more important.

We argue that these are the optical vortices appearing in numerical simulations of focused beams, and that the correct explanation is not SAM to OAM conversion but just the inherent spatial properties of light modes with definite energy, $P_z$, $J_z$ and $\Lambda$ propagating through a system that conserves energy, $\Lambda$ and $J_z$ while breaking transverse translational invariance. The lens shifts the weight distribution towards modes with larger $\krho$ values and optical vortices already existing in the initially strongly attenuated polarization components of the input beam gain relative importance in the focalized beam. 
For the theoretical study of optical vortices in focused beams, \cite{Zhao2007} and \cite{Bliokh2011} use solutions of the paraxial equation as the input modes, instead of using solutions of the full Maxwell equations as we have done. Since paraxial input modes do not have the attenuated phase singularities in the other polarization components because only a single polarization component is non zero, the appearance of optical vortices upon focusing was, contrary to this paper's explanation, attributed to SAM to OAM conversion.

It is interesting to note that the property associated with $\Lambda P_z/|\mathbf{P}|$ is not preserved upon focusing. Helicity is conserved but, due to the lack of of translational symmetry in the transverse plane, $P_\rho$ changes, which implies that $P_z$ changes. The fact that the lack of a ``geometrical'' symmetry breaks the conservation law corresponding to $\Lambda P_z/|\mathbf{P}|$ indicates that the symmetry transformation generated by such ``spin operator'' is, in general, not independent of the geometry of the problem. 

\subsection{Optical vortices in scattering}\label{sec:helvor}
The experimental observation of optical vortices in scattered fields has been reported in the literature \cite{Gorodetski2009,Vuong2010,Manni2011}. In these papers, the observations are explained by means of spin to orbit angular momentum conversion during the interaction with the scatterer. Recently, similar observations have been analyzed in \cite{FerCor2012} using symmetries and conserved quantities, and reaching a very different conclusion. In line with the discussion of the last paper, we will show in this section that the reason for all these experimental observations is not SAM to OAM transfer, but a partial helicity change during the light-matter interaction due to the breaking of electromagnetic duality symmetry in the system. We will also argue that the helicity change is enhanced by physical effects which strongly break duality, beyond the inherent duality asymmetry of general planar multilayer structures.

\begin{figure}[ht]
\begin{center}
\includegraphics[scale=1.25]{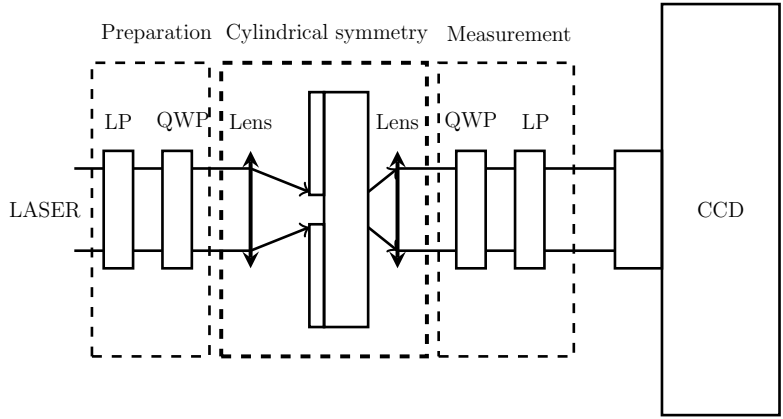}
\end{center}
\caption{Archetypal experimental setup. A collimated beam is circularly polarized by means of a linear polarizer (LP) and a quarter wave plate (QWP). After focusing, the beam interacts with a cylindrically symmetric target, in this example, a circular nano-aperture in a metallic film on a substrate. A portion of the scattered field is collected and collimated by a second lens, analyzed by a second set of QWP an LP, and its spatial profile is imaged into a Charged Couple Device (CCD) camera. The two orthogonal settings of the last LP allow the projection of the two collimated helicity modes (see the text for details).}
\label{Fig:telescope}
\end{figure}

We have already mentioned several times that helicity transforms independently of the geometry of the scatterer. In particular, it transforms independently of whether the considered system has cylindrical symmetry or not. Nevertheless, as seen below, a change in helicity is very clearly identifiable in the spatial patterns of the scattered fields when the system has cylindrical symmetry. Several of the experimental setups and input beams in the articles cited in this section have cylindrical symmetry and are similar to the system in Fig. \ref{Fig:telescope}, which we will now analyze. 

In Fig. \ref{Fig:telescope}, we distinguish the preparation and measurement apparatus from the focusing and scattering part of the setup. In the preparation stage, a collimated Gaussian beam goes through a linear polarizer (LP) and quarter wave plate (QWP), which results in a beam with well defined values of $J_z$ and $\Lambda$ (see more details below). In the central part, the beam is focused onto a cylindrically symmetric target and the resulting scattered light is collected by another lens. The measurement part of the setup uses another QWP and LP to project light with different helicity depending on the setting of the LP (see more details below). 

At first sight, the central part is cylindrically symmetric but lacks translational symmetry in the transverse plane. We should hence expect conservation of $J_z$ and, as in section (\ref{sec:focusing}), non-conservation of $P_\rho$. Since the lenses preserve helicity, the behavior of the system with respect to $\Lambda$ depends on whether the target sample meets condition (\ref{eq:epsmu}). Let us assume that it does not meet such condition. 

Let us imagine that, after the first LP and QWP, the input to the first lens is a predominantly RC polarized Gaussian beam. Assuming perfect optical elements, and using the same arguments as in section \ref{sec:focusing}, we can see that the input beam can be expanded using basis (\ref{eq:CD}) into modes with $\smallkrho$ and sharp values of $J_z$, $n=-1$, and of $\Lambda$, $\lambda=-1$. The first lens will focus the beam into the sample without altering either $J_z$ or $\Lambda$, then, since it is assumed that the sample does not meet (\ref{eq:epsmu}), part of the light will change helicity upon interacting with the cylindrically symmetric target and, after collimation with the second lens, there will be two types of collimated modes, both with $n=-1$, but differing in the value of helicity. Schematically representing the actions of the lenses and the cylindrically symmetric scatterer as transformations of fields expanded in the basis (\ref{eq:CD}) by coordinates $\cnkrho$ and $\dnkrho$, we may summarize the whole sequence as:

\begin{equation}
c_{-1,\smallkrho} \xrightarrow{\rm Lens} c_{-1,k_{\rho}} \xrightarrow{\rm Scattering} (c_{-1,k_{\rho}},d_{-1,k_{\rho}}) \xrightarrow{\rm Lens} \left(c_{-1,\smallkrho},d_{-1,\smallkrho}\right).
\end{equation}

The corresponding modal expressions for the components of the two types of output modes are:
\begin{equation}
\begin{split}
\mathbf{C}_{-1\smallkrho}(\rho,\theta,z)&\approx\sqrt{\frac{\krho}{\pi}}i^{2}\exp(i\kz z)J_{0}(\krho\rho)\rhat.\\
\mathbf{D}_{-1\smallkrho}(\rho,\theta,z)&\approx\sqrt{\frac{\krho}{\pi}}i^{2}\exp(i\kz z)J_{-2}(\krho\rho)\exp(-i2\theta)\lhat.
\end{split}
\end{equation}

That is, a predominantly RC beam similar to the input, without any phase singularity in $\rhat$ and a predominantly LC vortex beam with a singularity of charge two in $\lhat$. If we set the last LP to project the LC component of the output collimated beam, this vortex will appear in the CCD camera. 


We argue that these are the vortices observed in the cylindrically symmetric scattering experiments of \cite[Fig. 4]{Gorodetski2009}, \cite[Figs. 2(c)-(d)]{Vuong2010} and \cite{Manni2011}, and that the underlying reason for their appearance is not SAM to OAM transfer, but that the electromagnetic duality symmetry is broken in those systems.

The samples used in \cite{Gorodetski2009} and \cite{Vuong2010} consist of nano-apertures on metallic thin films. The breaking of duality in a general planar multilayer structure, and in particular when it includes a metal, is clear since the relative electric and magnetic constants of the system will not meet condition (\ref{eq:epsmu}). Nevertheless, as shown in \cite{FerCor2012}, the effect due to the multilayer alone is typically small in terms of helicity conversion. The reason why the helicity conversion is enhanced making it easily detectable, is that the nano-apertures allow for light to couple to the natural modes of the multilayer system, where duality is strongly broken. In a multilayer system, the natural modes are either TE or TM resonances. Since a mode with well defined helicity has equal contributions from TE and TM components (\ref{eq:MNAA}), a TE only or TM only resonance implies a strong breaking of electromagnetic duality. The interfaces between a metal and a dielectric allow the existence of surface plasmon polaritons of TM only character. As shown in \cite{FerCor2012}, the influence of these modes in the transmitted light is responsible for the experimentally detectable helicity change.

In \cite{Manni2011}, optical vortices are observed upon propagation of light through a planar semiconductor microcavity, still a cylindrically symmetric system as noted in that work. In this case, duality is strongly broken in the multilayer itself by the energy splitting between TE and TM modes in the structure. This splitting is identified in that work as the enabler for SAM to OAM conversion.

References \cite{Gorodetski2009} and \cite{Vuong2010} contain also results for non-cylindrically symmetric setups. Even though their study using $J_z$ is not as simple, we are confident that the generality of the methodology that we propose can rid their analysis from the artificial concept of SAM to OAM conversion, possibly leading to further physical insights. For instance, for the squared nano-apertures studied in \cite{Vuong2010}, the same argumentation about the nano-aperture mediated coupling onto strong duality breaking multilayer natural modes applies. We postulate that any instance where SAM to OAM conversion is invoked can be properly explained using our framework. 

\section{Conclusions and Discussion}\label{sec:conc}
Using the helicity of light fields to represent the polarization degrees of freedom, we have introduced a general framework for the study of light beams with angular momentum and their interactions with matter. In particular, the framework does not depend on the applicability of the paraxial approximation. Our proposal is based on two hermitian operators in the Hilbert space of transverse electromagnetic fields: the helicity $\Lambda$ and a component of angular momentum $J_z$. These operators are the generators of simple transformations of the fields: generalized electromagnetic duality and rotations along the $z$ axis respectively. Since the generalized duality transformation is independent from rotations or translations \cite{FerCor2012}, the use of $\Lambda$ allows to consider the changes in polarization independently from other degrees of freedom like angular momenta ($\mathbf{J}$) and linear momenta ($\mathbf{P}$). This simplification is crucial for the practical applicability of the framework. We recall that the approximations implicit in the macroscopic Maxwell's equations are assumed in this paper. 

We propose this new framework as a substitute of the current state of the art treatment, which is based on the separation of spin and orbital angular momenta for the description of the angular momentum of light, and on the mechanism of SAM to OAM transfer in light-matter interactions and other situations. We have shown that it avoids the fundamental problems associated with the separation of SAM and OAM in a simpler fashion than the current theoretical solutions. We have also demonstrated its predictive power based on analyzing the broken and unbroken symmetries of the system to predict which properties of the light will change and which ones will be preserved. The current framework lacks predictive power and, at this point, can only be qualified as a descriptive theory. This is illustrated by the fact that, using our methodology, we have shown that, phenomena commonly explained as SAM to OAM transfer in focusing and scattering are actually due to two completely independent physical reasons, showing the inconsistency of the current framework. The results of this paper suggest that such inconsistency may be related with the use of quantities like SAM and OAM, which are not valid operators for transverse Maxwell fields and hence cannot be related to meaningful transformations of the fields.

We are confident that the use of helicity and its associated transformation will become a powerful theoretical and practical tool which will improve the understanding and control of light-matter interactions at the nano-scale. Applications of the ideas presented in this paper can be foreseen in nano-optics, for the control of the shape and polarization of electromagnetic fields, in metrology, for probing the equivalence of the electric and magnetic responses of a system, and in quantum science and technology, where the identification of two truly independently observable degrees of freedom of the field like $\Lambda$ and $J_z$ should allow to better understand the potential benefits of the use of angular momentum in quantum applications.   

The association of helicity and duality with other degrees of freedom and their corresponding transformations constitutes a general and robust methodology to study practical light-matter interaction problems by using fundamental concepts.
\section*{Acknowledgements}
This work was funded by the Australian Research Council Discovery Project DP110103697 and the Center of Excellence in Engineered Quantum Systems (EQuS). G.M.-T is also funded by the Future Fellowship FF110100924.

\appendix
\section{Helicity eigenstates and transverse electric and transverse magnetic modes}\label{appA}
We set out to proof a general relationship between electromagnetic states of well defined helicity and the transverse electric and transverse magnetic modes. We start by making use of an elegant method for finding solutions of the monochromatic Maxwell equations in a source free, isotropic and homogeneous medium \cite[chap. 13.1]{Morse1953}, \cite[chap. VII]{Stratton1941}. Under some suitable conditions of the coordinate system $\left\lbrace u_1,u_2,u_3 \right\rbrace$, two transverse independent solutions of the vectorial Helmholtz equation of the medium can be obtained from a separable solution $\psi(u_1,u_2,u_3) = U_1(u_1)U_2(u_2)U_3(u_3)$ of the corresponding scalar Helmholtz equation. With $\mathbf{\hat{w}}$ a unit vector perpendicular to the surface of constant coordinate $u_1= C$, the two vector solutions are obtained as:
\begin{equation}
\label{eq:MN}
\Mr=\nabla \times (\mathbf{\hat{w}} \psi) \text{ and } \Nr=\frac{\nabla \times \Mr}{k},
\end{equation}
There are six different coordinate systems for which a complete basis for transverse electromagnetic fields can be built in this way \cite[chap. 13.1]{Morse1953}. Plane waves, multipoles and Bessel beams result from using cartesian, spherical and cylindrical coordinates, respectively \cite[chap. VII]{Stratton1941}. Besides being eigenvectors of the Hamiltonian due to its monochromatic character, each coordinate system produces electromagnetic modes which are eigenvectors of a different set of operators:
\begin{itemize}
\item plane waves with $\mathbf{\hat{w}}=\zhat$: eigenvectors the of transverse momenta $k_x$ and $k_y$.
\item Multipoles with $\mathbf{\hat{w}}=\frac{{\mathbf{r}}}{|{\mathbf{r}}|}$: eigenvectors of the squared angular momentum norm $J^2$ and a component of angular momentum, for instance $J_z$.
\item Bessel beams with $\mathbf{\hat{w}}=\zhat$: eigenvectors of the third components of the linear and angular momenta $P_z$ and $J_z$ .
\end{itemize}

From now on, we will lump the energy $H$, which is proportional to $k$, and these other degrees of freedom into a collective index $\nu$ and use $\Mreta$ and $\Nreta$. In all three reference systems, the $\Mreta$ and $\Nreta$ modes are commonly referred to as TE and TM modes, respectively.

Using $\nabla \times =\vec{\mathbf{S}}\cdot\vec{\mathbf{P}}$ \cite[XIII.93]{Messiah1958}, it can be proved that the helicity operator $\Lambda$ can be written in real space as $\Lambda=\frac{\nabla \times}{k}$. In such formulation, it is also true that $\Mreta=\Lambda \Nreta$ \cite[chap. 13.1]{Morse1953}. Then, together with (\ref{eq:MN}), we see that:
\begin{equation}
\Nreta = \Lambda \Mreta,\ \Mreta = \Lambda \Nreta. 
\label{hel_M_N}
\end{equation}
Namely, the TE and TM modes are transformed into each other by the application of $\Lambda$. We can hence obtain electromagnetic fields with well defined helicity as:
\begin{equation}
\label{eq:ApAm}
\begin{split}
\Aplus &= \frac{1}{\sqrt{2}}\left(\Mreta + \Nreta\right), \  \Lambda \Aplus = \Aplus\\
\Aminus &=\frac{1}{\sqrt{2}}\left(\Mreta - \Nreta\right), \  \Lambda \Aminus= -\Aminus.
\end{split}
\end{equation}
Equations (\ref{eq:ApAm}) already make it obvious that helicity conservation will only happen for equivalent scattering of the TE and TM components of the field, which is needed in order to preserve their linear combinations. 

It is clear that, since $\lbrace \Mreta,\Nreta \rbrace$ are an orthonormal basis for transverse fields when all values of $\nu$ are considered, so is $\lbrace \Aplus,\Aminus \rbrace$. This derivation of the TE, TM modes is valid for the systems of coordinates mentioned above. On the other hand, the helicity operator is well defined for any basis. Thus, we think that it is more natural to define the TE and TM modes as $\Mreta = \frac{1}{\sqrt{2}}\left(\Aplus + \Aminus\right),\ \Nreta = \frac{1}{\sqrt{2}}\left(\Aplus - \Aminus\right)$, and as such we use them in the main text.

\section{Derivation of Bessel beams with well defined helicity}\label{appB}
In this appendix we present a constructive derivation of electromagnetic modes with well defined energy $H$, third component of angular and linear momenta $J_z$ and $P_z$, and helicity $\Lambda$. 

We start by considering TE and TM modes of well defined energy and linear momentum. They are plane waves derived following the constructive procedure given in \cite[chap. 13.1]{Morse1953}-\cite[chap. VII]{Stratton1941} which was already used in section \ref{sec:scattering}. Their explicit expressions are, in the cartesian $[\xhat,\yhat,\zhat]$ basis:
\begin{eqnarray}
\label{eq:s}
\shat \cdot \plane & = &\frac{i}{k_{\rho}} \left( k_y \widehat{\textbf{x} } - k_x \widehat{\textbf{y} } \right) \exp(i \textbf{k} \cdot \textbf{r})\\
\label{eq:p}
\phat \cdot \plane & = &  \left[ \frac{ - \kz \left( k_x \xhat + k_y \yhat \right) + \krho^2 \zhat} { k \krho} \right] \plane,
\end{eqnarray}
where $\krho=\sqrt{k_x^2+k_y^2}$, and $[k_x,k_y,k_z]$ and $k$ are proportional to the linear momentum vector and energy of the plane waves. As per equation (\ref{eq:ApAm}), sum and subtraction of TE and TM modes result in states of well defined helicity:
\begin{equation}
\begin{split}
\pwplus\plane&=\frac{1}{\sqrt{2}}\left(\pws+\pwp\right)\plane,\\
\pwminus\plane&=\frac{1}{\sqrt{2}}\left(\pws-\pwp\right)\plane.
\end{split}
\end{equation}

Interestingly, any plane wave of well defined helicity and momentum vector proportional to $[k_x,k_y,k_z]$ can also be obtained \cite{Gabi2008} by rotating a plane wave of the same helicity and momentum $[0,0,k]$:
\begin{equation}
\label{eq:rothel}
\begin{split}
\pwplus\plane&=\rotation\left(-\lhat\exp(ikz)\right)\\
\pwminus\plane&=\rotation\left(\rhat\exp(ikz)\right).
\end{split}
\end{equation}
where $\lhat=\frac{\xhat + i \yhat}{\sqrt{2}}$, $\rhat=\frac{\xhat - i \yhat}{\sqrt{2}}$, $\theta_k=\arcsin{\frac{\krho}{k}}$ and $\phi_k=\arctan{\frac{k_y}{k_x}}$. 
Equation (\ref{eq:rothel}) exploits the fact that helicity does not change under spatial rotations.

A rotation operation of a vectorial field $\rotation \left(\mathbf{A}(\mathbf{r})\right)$ is, explicitly:
\begin{equation}
\mathbf{A}(\mathbf{r})\rightarrow R(\phik,\thetak)\mathbf{A}(R^{-1}(\phik,\thetak)\mathbf{r}),
\end{equation}
where $R(\phik,\thetak)$ is the rotation matrix:
\begin{equation}
R(\phik,\thetak)=R_z(\phik)R_y(\thetak)= \begin{pmatrix} \ct \cp & - \spp & \st \cp \\  \ct \spp & \cp & \st \spp \\ -\st & 0 & \ct \end{pmatrix},
\end{equation}
where $R_z(\phik)$ and $R_y(\thetak)$ are rotations around the $\zhat$ and $\yhat$ axis respectively.

We now start our construction of modes with well defined $H$, $P_z$, $J_z$ and $\Lambda$ by expressing the most general monochromatic forward propagating transverse electromagnetic field as a combination of plane waves with well defined momentum:
\begin{equation}
\label{eq:A}
\mathbf{A}= \int_0^{\frac{\pi}{2}} sin \thetak d\thetak \int_0^{2\pi} d\phik \left[ \alphak \rotation \left(-\lhat \exp(ikz)\right) + \betak \rotation \left(\rhat \exp(ikz)\right) \right].
\end{equation}
A backward propagating beam can be obtained by using $-\lhat \exp(-ikz)$ and $\rhat \exp(-ikz)$ as the initial plane waves instead.

Having restricted (\ref{eq:A}) to a single wavenumber $k$ assures that the resulting mode has a definite energy $H$ proportional to $k$. With respect to $\Lambda$, it is clear from (\ref{eq:A}) that, if we desire a field $\mathbf{A}$ with well defined helicity, we need only to set either $\alphak$ or $\betak$ equal to zero for all $(\theta_k,\phi_k)$, so that only plane waves of the same helicity type are present in (\ref{eq:A}). With respect to having a well defined $P_z$, and because, in a plane wave, its value is proportional to $k_z=k\cos\thetak$, we must include only a single value of $\theta_k$ in (\ref{eq:A}). 

The conditions needed in (\ref{eq:A}) to specify a beam with well defined $J_z$ are not as apparent as in the case of $H$, $P_z$ and $\Lambda$. The solution can be reached by applying the operator $J_z$ to the general mode $\mathbf{A}$. In order to do that, we use the following definition of $J_z$ as an infinitesimal rotation operation:
\begin{equation}
\label{eq:jz}
J_z=\lim_{d\phik\rightarrow 0} \frac{I-\mathbf{R}(0,d\phik)}{id\phik},
\end{equation}
where $I$ is the identity operator.

Applying (\ref{eq:jz}) to (\ref{eq:A}) and making use of the properties of rotation operators we obtain:
\begin{equation}
\begin{split}
J_z [\mathbf{A}]  =\lim_{d\phik\rightarrow 0} \int_0^{\frac{\pi}{2}} \sin \thetak d \thetak \int_0^{2\pi} d\phik \alphak \frac{\rotation - \mathbf{R}(\thetak,\phik + d\phik) }{id\phik} \left(-\lhat \exp(ikz)\right) +\\  \betak \frac{\rotation - \mathbf{R}(\thetak,\phik + d\phik) }{id\phik} \left(\rhat \exp(ikz)\right),
\end{split}
\end{equation}
which, after integrating by parts can be reduced to:
\begin{equation}
J_z [\mathbf{A}]  = \int_0^{\frac{\pi}{2}} \sin \thetak d \thetak \int_0^{2\pi} d\phik \frac{\partial \alphak}{i \partial \phik} \rotation \left(-\lhat \exp(ikz)\right)+\frac{\partial \betak}{i \partial \phik} \rotation \left( \rhat \exp(ikz)\right).
\end{equation}
So, in this representation, the operator $J_z$ acts by taking the partial derivative of the coordinates $(\alphak,\betak)$ with respect to $\phik$ and dividing by $i$.

Gathering together all the above, we can assert that 
\begin{equation}
\alphak = \frac{1}{2\pi} \delta(\thetak - \thetak')\exp{(in\phik)},\  \betak=0,
\end{equation}
specifies a mode with well defined energy $H$ proportional to $k$, well defined $\Lambda$ with value $\lambda=1$, well defined $J_z$ with value $n$ and well defined $P_z$ proportional to $k_z=k\cos(\thetak')$, and that
\begin{equation}
\alphak = 0 ,\  \betak=\frac{1}{2\pi} \delta(\thetak - \thetak')\exp{(in\phik)},
\end{equation}
does the same for the opposite helicity.
 
After inserting the specified coordinates into equation (\ref{eq:A}), substituting the rotated plane wave by their explicit expressions as linear combinations of (\ref{eq:s}) and (\ref{eq:p}), changing basis from $[\xhat,\yhat,\zhat]$ to $[\rhat,\lhat,\zhat]$ and using $\plane = \sum_m i^m J_m(\krho \rho ) \exp(i m (\phi - \phik)) \exp(i\kz z)$ before performing the integral in $d\phik$, we finally obtain the real space expressions of the modes in cylindrical coordinates that have been extensively used in the main text:

{\scriptsize
\begin{equation}
\begin{split}
\label{eq:CD2}
\Cnkrho(\rho,\phi,z)&=\sqrt{\frac{\krho}{2\pi}}i^n\exp(i(\kz z+n\phi))\left(\frac{i}{\sqrt{2}}\left((1+\frac{\kz}{k})J_{n+1}(\krho\rho)\exp(i\phi)\rhat+(1-\frac{\kz}{k})J_{n-1}(\krho\rho)\exp(-i\phi)\lhat \right)-\frac{\krho}{k}J_n(\krho\rho)\zhat\right),\\
\Dnkrho(\rho,\phi,z)&=\sqrt{\frac{\krho}{2\pi}}i^n\exp(i(\kz z+n\phi))\left(\frac{i}{\sqrt{2}}\left((1-\frac{\kz}{k})J_{n+1}(\krho\rho)\exp(i\phi)\rhat+(1+\frac{\kz}{k})J_{n-1}(\krho\rho)\exp(-i\phi)\lhat \right)+\frac{\krho}{k}J_n(\krho\rho)\zhat\right).\\
\end{split}
\end{equation}
}

Note that, due to the integration limits of $\thetak$ in (\ref{eq:A}), the derivation is restricted to propagating modes. In reality, values of $\krho>k$ in equations (\ref{eq:CD2}) are possible, specifying non-propagating modes with well defined $H$, $P_z$, $J_z$ and $\Lambda$. Including non-propagating modes from the start of the derivation can be done by using imaginary values of $\thetak$ in order to obtain values of $\sin(\theta)$ bigger than one.

\section{Preservation of $J_z$ and $\Lambda$ by the action of an aplanatic lens}\label{appC}

The aplanatic or spherical lens model allows to relate the real space field profile of the collimated input beam to the angular spectrum of the focused beam. Originally developed by Richards and Wolf in \cite{Richards1959}, we can find an explanation of the model and its formulas in \cite[chap. 3.5]{Novotny2006}, which we reproduce here using a slightly different notation using that $\exp(ikz \cos \thetak) \exp (ik \rho \sin \thetak \cos (\phi - \phik))=\plane$:

{\scriptsize
\begin{align}
\label{eq:aplanatic_us}
& \mathbf{E}_{out}(\rho,\phi,z)  =  \frac{i k f \exp (-ikf)}{2\pi} \int_0^{\thetak^{m}} \int_0^{2\pi}  \sin \thetak  d \thetak d\phik \ \mathbf{E}_{\infty}(\thetak,\phik) \plane\\
\label{eq:aplanatic_inc}
& \Einf  = \left[  t^s(\thetak)  \left(\pwsthetaphi{0}{\phik} \cdot  \mathbf{E}_{inc}(f\sin(\thetak),\phik)  \right) \pwsthetaphi{\thetak}{\phik} + t^p(\thetak) \left( \pwpthetaphi{0}{\phik} \cdot \mathbf{E}_{inc}(f\sin(\thetak),\phik)  \right) \ \pwpthetaphi{\thetak}{\phik}    \right] \sqrt{\frac{n_1}{n_2}}(\cos \thetak)^{1/2}.
\end{align}
}

Where:
\begin{itemize}
\item $\mathbf{E}_{out}(\rho,\phi,z)$ is the focused field in real space expressed in cylindrical coordinates $[\rho,\phi,z]$.
\item $f$ is the focal distance of the lens, $\thetak^{m}=\arcsin(\textrm{NA})$, where $\textrm{NA}$ is the numerical aperture of the lens, and $k$ the wavenumber of the field.
\item $[\rho,\phi,z=z_0]$, with $\rho=f\sin(\thetak)$ and $\phi=\phik$  are both the real space cylindrical coordinate system for the input beam $\mathbf{E}_{inc}(\rho,\phi,z=z_0)$ and the spherical coordinates in momentum space of $\Einf$, the angular spectrum of the focalized output beam, with $\thetak=\arcsin(\frac{\krho}{k}),\phik=\arctan(\frac{k_y}{k_x})$. This dual role of the coordinates is the essence of the model.
\item $(t^s(\thetak),t^p(\thetak))$ are the lenses TE and TM transmission coefficients and $(n_1,n_2)$ the indexes of refraction of the input and output media.
\end{itemize}

We emphasize that the model is valid for a collimated input only. 

The definitions of the polarization vectors $\pwsthetaphi{\alpha}{\psi}$ and $\pwpthetaphi{\alpha}{\psi}$ are those of (\ref{eq:s}) and (\ref{eq:p}) with $\alpha=\arccos(\frac{k_z}{k})$ and $\psi=\arctan(\frac{k_y}{k_x})$.

We now analyze the properties of the model with respect to conservation of $J_z$ and $\Lambda$. 
We start by using some of the ideas and techniques of appendix \ref{appB} in order to calculate the angular momentum of the focused beam when the input beam has a definite value of $J_z$.

Using (\ref{eq:A}), expression (\ref{eq:aplanatic_us}) can also be written as:
{\scriptsize
\begin{equation}
\label{eq:sum}
\mathbf{E}_{out}(\rho,\phi,z)= \int_0^{\pi} \sin (\thetak) d\thetak \int_0^{2\pi} d\phi\left[ g_s(\thetak, \phik )\rotation \left(-i\yhat\plane\right) + g_p(\thetak, \phik ) \rotation \left(-\xhat\plane\right)\right],
\end{equation}
}
where
\begin{equation}
\label{eq:gsgp}
\begin{split}
g_s(\thetak, \phik )&= t^s(\thetak) \left(\pwsthetaphi{0}{\phik}\cdot \mathbf{E}_{inc}( f\sin(\thetak),\phik) \right)\\
g_p(\thetak, \phik )&= t^p(\thetak) \left(\pwpthetaphi{0}{\phik}\cdot \mathbf{E}_{inc}( f\sin(\thetak),\phik) \right).
\end{split}
\end{equation}
As seen in appendix \ref{appB}, in order to apply the operator $J_z$ to the output focused beam in (\ref{eq:sum}), we need to take the partial derivative of its coordinates $(g_s(\thetak,\phik),g_p(\thetak,\phik))$ with respect to $\phik$ and divide by $i$. Equation (\ref{eq:gsgp}) relates $(g_s(\thetak,\phik),g_p(\thetak,\phik))$ to the input beam. Note that the effect of the lens, which is given by the transmittivities $t^s(\thetak),t^p(\thetak)$, does not add any azimuthal dependence to $g_s(\thetak, \phik )$ and $g_p(\thetak, \phik )$. Thus, with this model, the lens will not affect the azimuthal dependence of the input beam, hence keeping the angular momentum constant. More explicitely, since vectors $\pwsthetaphi{0}{\phik},\pwpthetaphi{0}{\phik}$ are proportional to the azimuthal and radial polarization vectors defined as:
\begin{equation}
\phihat=\begin{pmatrix} -\sin\phik\\\cos\phik\\0\end{pmatrix},\
\rhohat=\begin{pmatrix} \cos\phik\\\sin\phik\\0\end{pmatrix},\
\end{equation}
the focalized field angular spectrum coordinates at point $(\thetak, \phik)$, $(g_s(\thetak, \phik ),g_p(\thetak, \phik ))$, are the real space azimuthal $E_{inc}^\phihat$ and radial $E_{inc}^\rhohat$ components of the input field at point $\rho=f\sin(\thetak),\phi=\phik$. In order to compute the angular momentum of the focused field we need to take the partial derivatives of the radial and azimuthal components of the input field with respect to $\phik=\phi$ and divide by $i$. Let us now find explicit expressions for $E_{inc}^\phihat$ and $E_{inc}^\rhohat$.

The collimated input beam can be expanded in the plane $z=z_0$ into modes of the type (\ref{eq:Cparaxial}) and  (\ref{eq:Dparaxial}) with $\smallkrho$, which we rewrite here after the change of basis $\rhat=\frac{1}{\sqrt{2}}(\rhohat-i\phihat)\exp(-i\phi)$, $\lhat=\frac{1}{\sqrt{2}}(\rhohat+i\phihat)\exp(i\phi)$:

\begin{equation}
\label{eq:cd}
\begin{split}
\Cnkrho(\rho,\phi,z=z_0)&\approx\sqrt{\frac{\krho}{\pi}}i^{n+1}\exp(i\kz z_0)J_{n+1}(\krho\rho)\exp(i\phi n)\frac{1}{\sqrt{2}}(\rhohat+i\phihat),\\
\Dnkrho(\rho,\phi,z=z_0)&\approx\sqrt{\frac{\krho}{\pi}}i^{n+1}\exp(i\kz z_0)J_{n-1}(\krho\rho)\exp(i\phi n)\frac{1}{\sqrt{2}}(\rhohat-i\phihat).
\end{split}
\end{equation}

Note that in order to avoid confusion with $\thetak$, the letter for the input beam coordinate angle $\arctan(y,x)$ is now $\phi$, instead of the original letter $\theta$ in expressions (\ref{eq:Cparaxial}) and (\ref{eq:Dparaxial}). 

The sum and subtraction of the above modes result in approximately pure radially and azimuthally polarized modes, which separates the radial and azimuthal components of the field:
\begin{equation}
\label{eq:ab}
\begin{split}
\Ankrho(\rho,\phi,z=z_0)&\approx\sqrt{\frac{2\krho}{\pi}}i^{n+1}\exp(i\kz z_0)\left[J_{n+1}(\krho\rho)+J_{n+1}(\krho\rho)\right]\exp(i\phi n)\frac{1}{\sqrt{2}}\rhohat=A^\rhohat_{n\krho}(\rho,\phi,z_0) \rhohat,\\
\Bnkrho(\rho,\phi,z=z_0)&\approx-\sqrt{\frac{2\krho}{\pi}}i^{n+2}\exp(i\kz z_0)\left[J_{n+1}(\krho\rho)-J_{n+1}(\krho\rho)\right]\exp(i\phi n)\frac{1}{\sqrt{2}}\phihat=B^\phihat_{n\krho}(\rho,\phi,z_0) \phihat,\\
\end{split}
\end{equation}
which means that, (\ref{eq:gsgp}) is explicitly
\begin{equation}
\begin{split}
g_s(\thetak, \phik )&= t^s(\thetak)E_{inc}^\phihat=t^s(\thetak)B^\phihat_{n\krho}(f\sin(\thetak),\phik,z_0), \\
g_p(\thetak, \phik )&= t^p(\thetak)E_{inc}^\rhohat=t^p(\thetak)A^\rhohat_{n\krho}(f\sin(\thetak),\phik,z_0).
\end{split}
\end{equation}

Then, we need to take the partial derivative of the azimuthal and radial input field components $B^\phihat_{n\krho}(f\sin(\thetak),\phik,z=z_0)$ and $A^\rhohat_{n\krho}(f\sin(\thetak),\phik,z=z_0)$ with respect to $\phik$ and divide by $i$. Let us now assume that the input is a field with definite angular momentum $J_z=n$. Inspection of (\ref{eq:ab}) reveals that such operation will leave those components invariant except for a multiplicative factor equal to $n$. So, the output field is also a field with definite angular momentum equal to $n$. We have just proved that the aplanatic lens model transfers the $J_z$ value of the input beam to the output beam without changing it.

We now study the behavior of the model with respect to helicity. From (\ref{eq:gsgp}), it becomes clear that the TE and TM content of the output field are independently determined by the real azimuthal and radial components (\ref{eq:ab}) of the input field respectively. Having taken linear combinations of modes with defined helicity, and using the results in appendix \ref{appA}, it follows that the modes in (\ref{eq:ab}) are also pure TE and TM modes respectively. Using the results from section \ref{sec:scattering} about helicity preservation in the TE TM basis, we now see that, since there is no cross-talk between the TE and TM components of the input and focalized fields, the key condition for helicity preservation in the aplanatic lens model is:

\begin{equation}
t_s(\thetak)=t_p(\thetak) \ \forall \ \thetak.
\end{equation}

In real manufacturing of microscope objectives, this condition is related to the coating of the lens surfaces \cite[chap. 3.6]{Novotny2006}, that is, a property of the materials and not the geometry of the system.

Let us assume for a moment a more general dependence of the transmission coefficients $t_s(\thetak,\phik),t_p(\thetak,\phik)$. The new $\phik$ dependence could destroy the $J_z$ preservation since now more terms will be involved in the partial derivatives of $g_s(\thetak, \phik )$ and $g_p(\thetak, \phik )$. On the other hand, as long as $g_s(\thetak, \phik )=g_p(\thetak, \phik )$ for all $(\thetak,\phik)$, helicity will be preserved. This is another example of the independence of the conservation laws of $J_z$ and $\Lambda$.


\end{document}